\RequirePackage{lineno}
\documentclass{webofc}
\usepackage[varg]{txfonts}   
\usepackage[varg]{xcolor}   
\usepackage{amsmath,epsfig}
\usepackage{graphicx}
\usepackage{graphics}
\wocname{EPJ Web of Conferences}
\woctitle{Resonance Workshop at Catania}
\begin{document}
%


\title{$K^{*0}$(892) and $\phi$(1020) resonance production at RHIC}
%
%

\author{Lokesh Kumar \inst{1} 
 \fnsep\thanks{\email{lokesh@rcf.rhic.bnl.gov}}
(for STAR Collaboration)
}

\institute{Department of Physics, Panjab University, Chandigarh, INDIA-160014 
          }

\abstract{%
The measurement of resonance production in ultrarelativistic heavy-ion 
collisions provides a glimpse of the hadronic medium properties and
its evolution at different stages. Resonances decaying into hadrons are 
used to estimate the time span and hadronic interaction cross section
in the hadronic phase between chemical and kinetic
freeze-out. 
Specifically, the comparison of $K^{*0}$(892) and
$\phi$(1020) resonances is interesting as the lifetimes of these
particles differ by about a factor of
10. 
Moreover, the
nuclear modification factor and azimuthal anisotropy measurements of 
mesonic resonances, which measure parton energy loss in medium and 
reflect partonic collectivity, can also probe particle-species and mass ordering.

The $K^{*0}$(892) and $\phi$(1020) resonance production at mid-rapidity
($|y|<$0.5), measured in high energy (Au+Au, Cu+Cu, d+Au
and $p+p$) collisions at RHIC with the STAR experiment, reconstructed
by their hadronic decay in $K\pi$ and $KK$, respectively, are
discussed.  Mesons' spectra, yields, 
mean transverse 
momentum $\langle p_T \rangle$, nuclear modification factor,
and azimuthal anisotropy are discussed as a function of centrality and
collision energy. 
}
\maketitle
\section{Introduction}
\label{intro}
Quantum Chromodynamics (QCD) predicts a phase transition from nuclear
matter to a state of deconfined matter, called the quark gluon plasma (QGP),
at high temperature and 
energy density~\cite{Karsch:2001vs}. High energy
heavy-ion collisions provide ideal scenario for the formation of the QGP~\cite{Adams:2005dq}.
The study of resonance production can be a useful tool in
understanding the properties of the system formed in heavy-ion collisions. 
Since their lifetimes are comparable to that of fireball, resonance particles are expected to
decay, rescatter, and regenerate to the kinetic freeze-out state
(vanishing elastic collisions). As a
result, their characteristic properties 
may be modified due to in-medium effects~\cite{medium}. The study of
resonance particles can provide the
time span and hadronic interaction cross-section of the hadronic phase
between chemical (vanishing inelastic collisions) and kinetic
freeze-out~\cite{time_span}.

Comparisons of $K^{*0}$ and $\phi$ mesons are interesting since their lifetimes
differ by a factor of 10~\cite{pdg}. $K^{*0}$ has a lifetime $\sim$4
fm/$c$~\cite{rapp}, comparable with that of the fireball, so it is expected to
suffer changes from the in-medium effects. There are two
competing processes affecting $K^{*0}$ yield, rescattering that
reduces the $K^{*0}$ yield and regeneration that may lead to increase in
$K^{*0}$ yield~\cite{rescatter_regen}. As a result, it is expected that $K^{*0}$/$K$ should
change as a function of centrality, i.e., increase due to regeneration or
decrease due to rescattering. On the other hand, $\phi$ mesons are
expected to freeze-out early~\cite{Adams:2005dq} and have a comparatively larger lifetime
($\sim$45 fm/$c$)
than $K^{*0}$, so they may not undergo rescattering and regeneration
effects. As a result, the $\phi$/$K$ ratio should remain constant as a
function of centrality. 

Another interesting property of the $\phi$ meson is that being a meson,
it has a mass comparable to that of protons and $\Lambda$, which are
baryons. Studying the $\phi$ meson (e.g., elliptic flow) along with these baryons and 
other mesons may give information on quark coalescence or the partonic
phase at the top RHIC energy~\cite{quark_coal}. Once the partonic phase is established at
higher energies, one would expect the turn-off of partonic phase or
decrease of dominance of the partonic interactions when the energy is
decreased. This is one of the goals of the RHIC Beam Energy Scan (BES) program~\cite{bes}.

\section{Experimental Detail}
\label{sec-1}
The results presented here are mainly from the Solenoidal Tracker
At RHIC (STAR) experiment. The STAR detector has a coverage of 2$\pi$ in azimuth and
pseudorapidity $|\eta|<$1. The data sets include Au+Au, Cu+Cu, $d$+Au, and $p+p$
collisions for energies $\sqrt{s_{NN}}=$ 62.4 and 200 GeV. Results from the Beam
Energy Scan phase-I, that include data from Au+Au collisions at
$\sqrt{s_{NN}}=$7.7, 11.5, 19.6, 27, and 39 GeV, are also
presented. The STAR Time Projection Chamber (TPC) is the main detector used for particle
identification 
by measuring the particle energy loss~\cite{tpc}. 

The centrality selection is done using the
uncorrected charged track multiplicity measured in the TPC within $|\eta|<$0.5 and
comparing with Monte-Carlo Glauber simulations~\cite{centrality}.
Both $K^{*0}$ and $\phi$ resonances are reconstructed via their
hadronic decay channels: $K^{*0} \rightarrow K\pi$ and $\phi
\rightarrow KK$. 
These daughter particles are identified using the TPC as mentioned above.
$K^{*0}$ and $\phi$ mesons are reconstructed by calculating invariant
mass for each unlike-sign  $K\pi$ and  $KK$, respectively, in an
event. 
The resultant distribution consists of the true signal ($K^{*0}$ or $\phi$) and
contributions arising from the random combination of unlike sign
$K\pi$ and  $KK$ pairs. To extract the $K^{*0}$ or $\phi$ yield, the
large random combinatorial background must be subtracted from the 
unlike sign $K\pi$ or $KK$ pairs. The random combinatorial background
distribution is obtained using the mixed-event technique~\cite{mix_event}.
In the mixed event technique, the background distribution is built
with uncorrelated unlike-sign $K\pi$ or $KK$ pairs from different
events. The generated mixed events distribution is then properly
normalized to subtract the background from the same event unlike-sign
invariant mass spectrum. 

\section{Results and Discussions}
\begin{figure*}
\centering
\sidecaption
 \includegraphics[height=5cm,width=6cm]{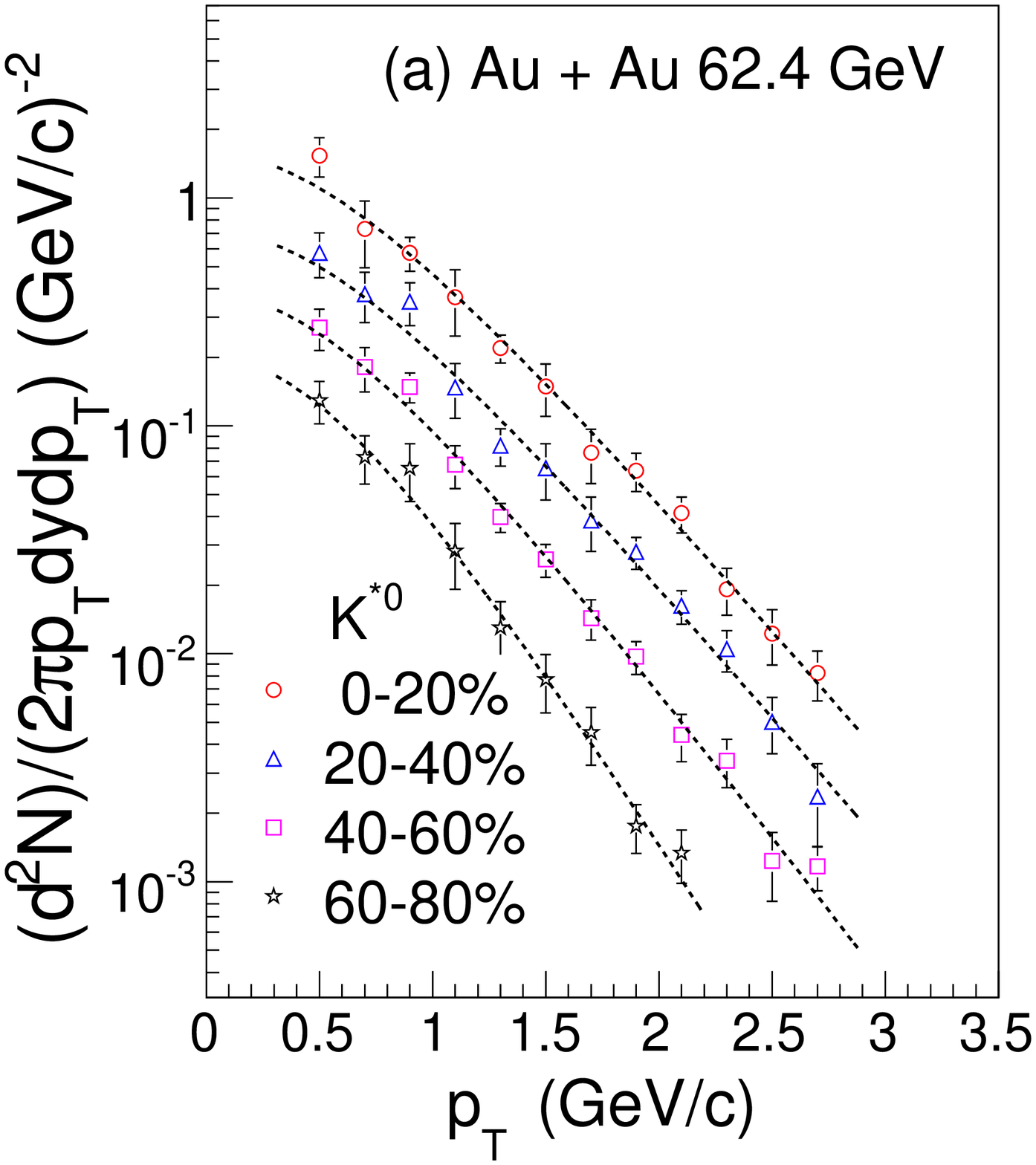}
 \includegraphics[height=5cm,width=5cm]{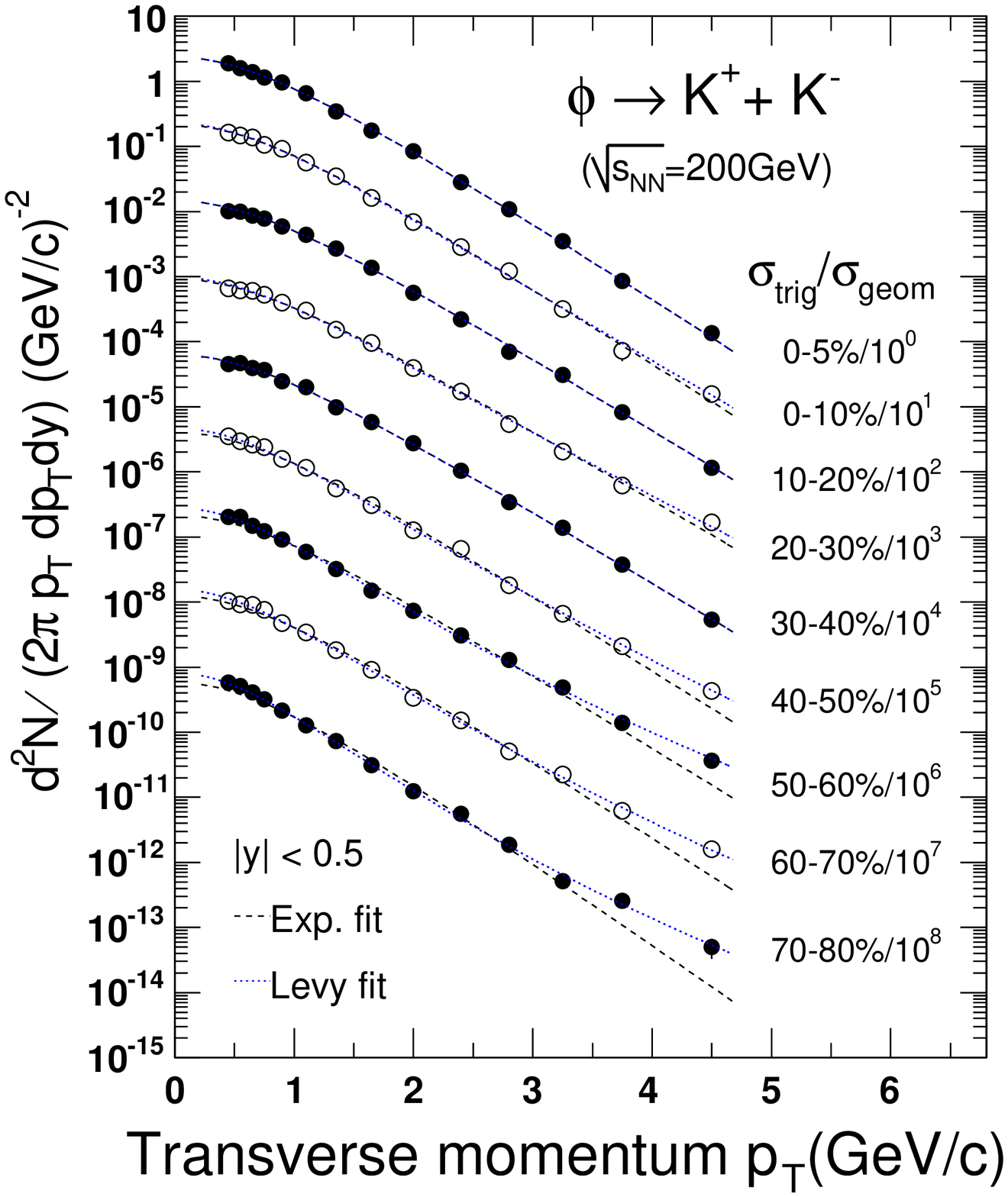}
\vspace{-2cm}
\caption{Left: Invariant yields versus $p_T$ for $K^{*0}$ in Au+Au
  collisions at 62.4 GeV for different centralities~\cite{Aggarwal:2010mt}. Right: Invariant
  yields versus $p_T$ for $\phi$ mesons in Au+Au collisions at 200 GeV
  for different centralities~\cite{Abelev:2007rw}.}
\label{spectra}       
\end{figure*}

Figure~\ref{spectra} shows the invariant yields versus transverse
momentum $p_T$ of $K^{*0}$ (left plot) and
$\phi$ (right plot) in Au+Au collisions at 62.4 and 200 GeV,
respectively for different collision centralities~\cite{Aggarwal:2010mt,Abelev:2007rw}. From these
distributions, $dN/dy$ and average transverse momentum $\langle p_T \rangle$ can be 
obtained. These quantities provide important information about the
system formed in high energy collisions.
It is observed that $dN/dy$ per participating nucleon pair for $K^{*0}$ increases
with increasing energy. 
$\phi$ meson yields per participating nucleon increases with
increasing energy and centrality~\cite{Aggarwal:2010mt,Abelev:2008aa}. $dN/dy$ per participating nucleon pair 
increases from $pp$, $d$+Au to Au+Au collisions at 200 GeV for both
$K^{*0}$ and $\phi$ mesons.
Comparing $\langle p_T \rangle$ of $K^{*0}$ with pions, kaons, and
protons in Au+Au collisions at 200 GeV, suggests that it is greater than that of pions and kaons but similar 
to that of protons, reflecting mass dependence or collectivity~\cite{kstar_meanpt}. When $K^{*0}$ $\langle p_T \rangle$ 
is compared between $pp$ and Au+Au collisions at 200 GeV, 
it is found to be larger
in Au+Au collisions, suggesting larger radial flow.
In general,  $\langle p_T \rangle$ increases with
particle mass, showing a collective behavior~\cite{Abelev:2008aa}.
\begin{figure}
\centering
\sidecaption
\includegraphics[height=4.2cm,width=6.cm]{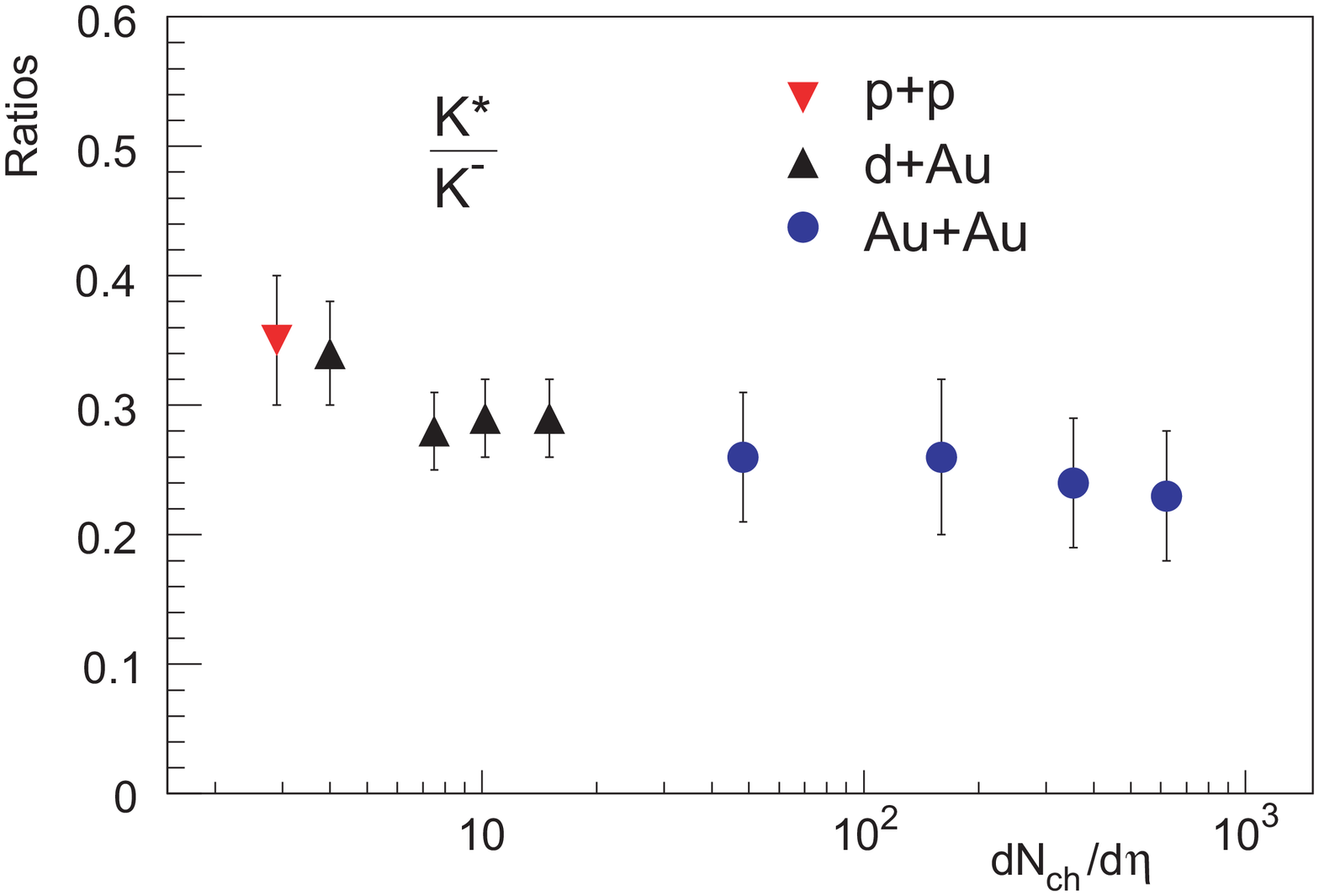}
\includegraphics[width=0.4\linewidth,trim=1 1 1 1,clip]{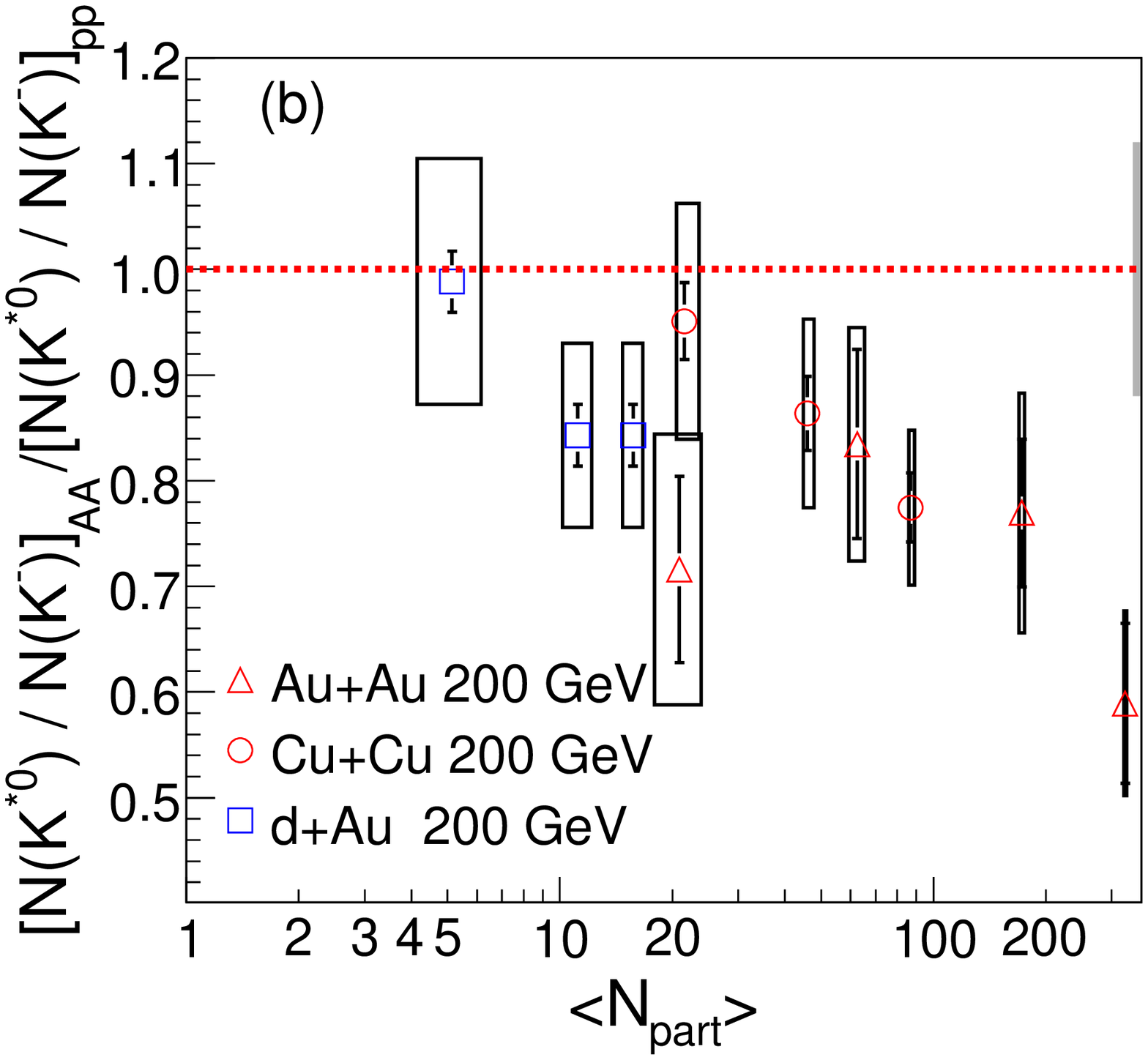}
\vspace{-4.0cm}
\caption{Left: $K^{*0}$/$K^-$ ratio in $p+p$ and various centralities
  in $d$+Au and Au+Au collisions as a function of
  $dN_{\rm{ch}}/d\eta$~\cite{Abelev:2008yz}.
Right: $K^{*0}$/$K^-$ ratio in Au+Au, Cu+Cu, and d+Au collisions
divided by that in $p+p$ collisions at $\sqrt{s_{NN}}=$ 200 GeV as a
function of $\langle N_{\rm{part}} \rangle$~\cite{Aggarwal:2010mt}.
 }
\label{ratio_kstar}       
\end{figure}
\begin{figure}
\includegraphics[scale=0.45]{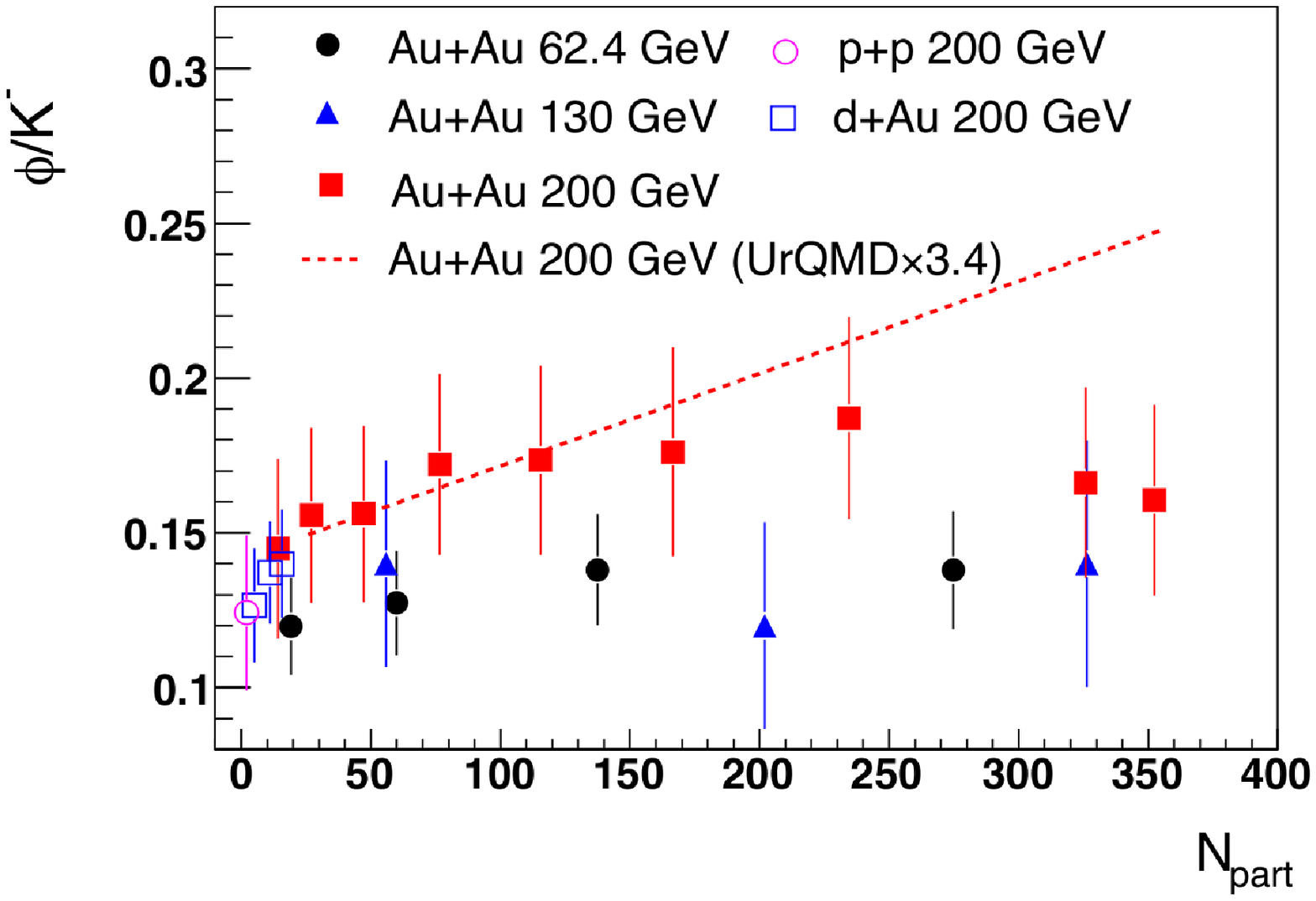}
\includegraphics[width=0.45\linewidth,trim=1 1 1 1,clip]{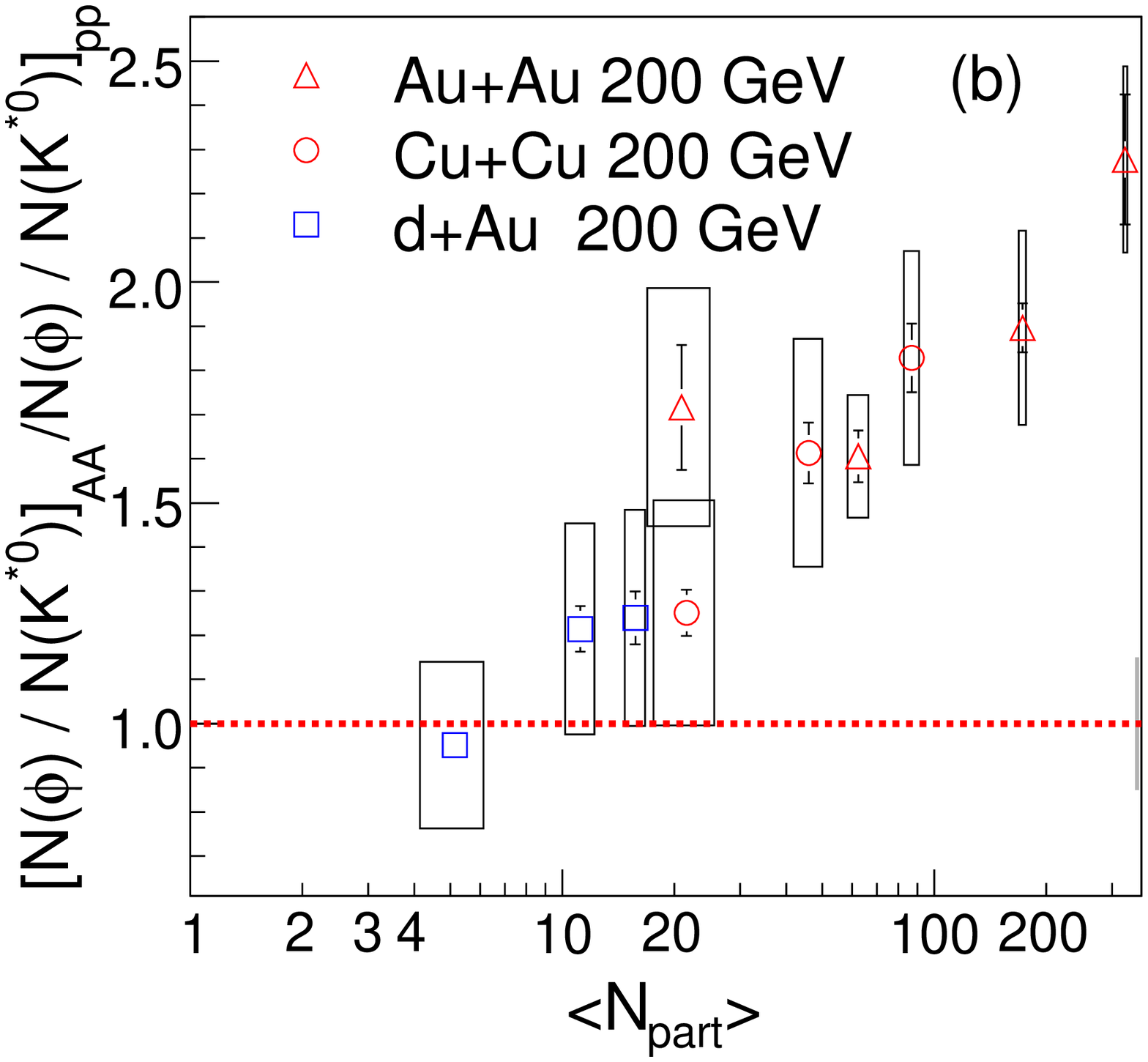}
\caption{Left: $\phi/K^{-}$ ratio in $p+p$ and various centralities
in $d$+Au and Au+Au collisions as a function of
$\langle N_{\rm{part}} \rangle$~\cite{Abelev:2008aa}. The dashed line shows results from UrQMD model calculations.
Right: $\phi/K^{*0}$ ratio in Au+Au, Cu+Cu, and d+Au collisions
divided by that in $p+p$ collisions at $\sqrt{s_{NN}}=$ 200 GeV as a
function of $\langle N_{\rm{part}} \rangle$~\cite{Abelev:2008yz}.}
\label{ratio_phi}       
\end{figure}
\begin{figure}
\centering
\sidecaption
\includegraphics[scale=0.3]{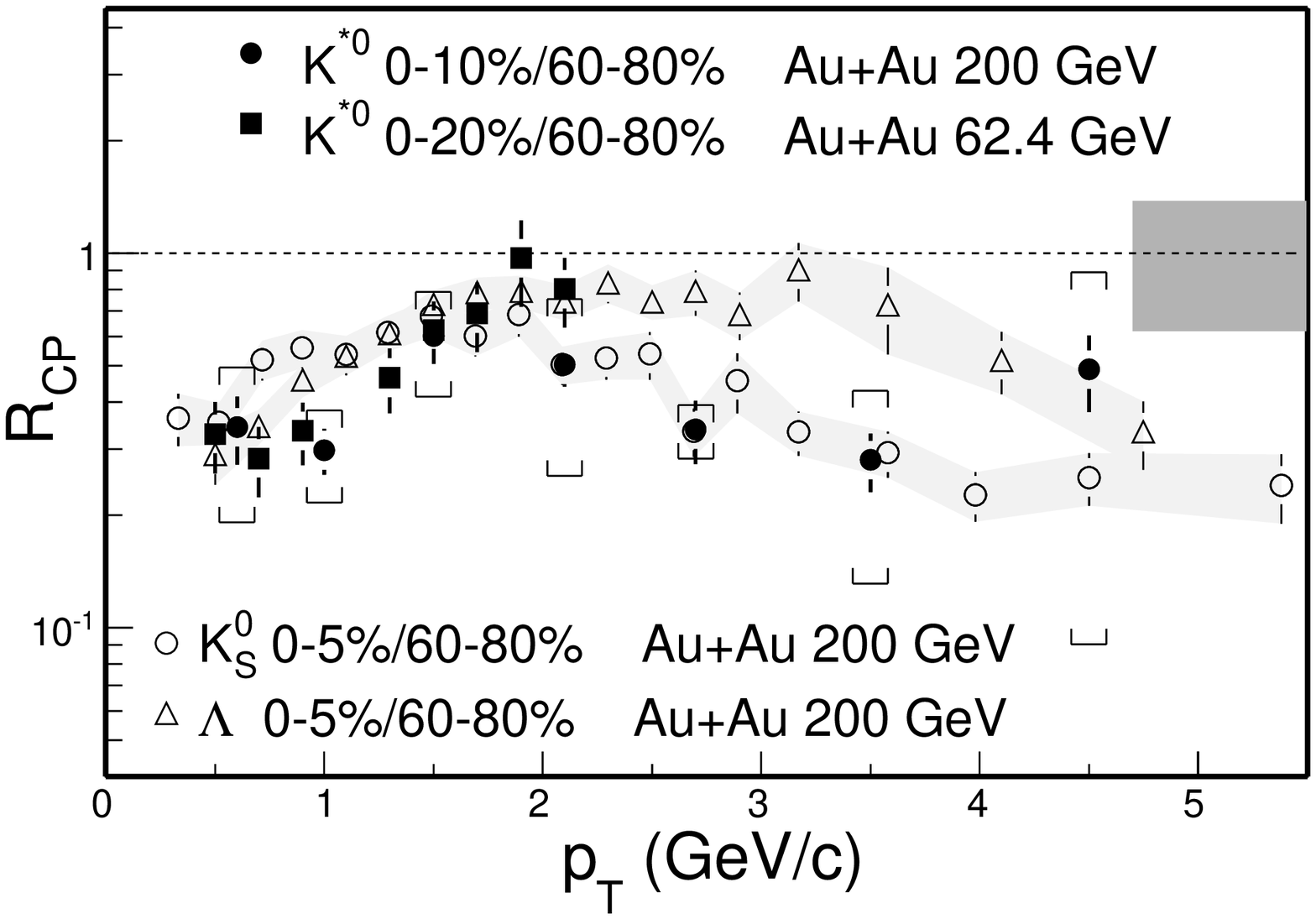}
 \includegraphics[scale=0.33]{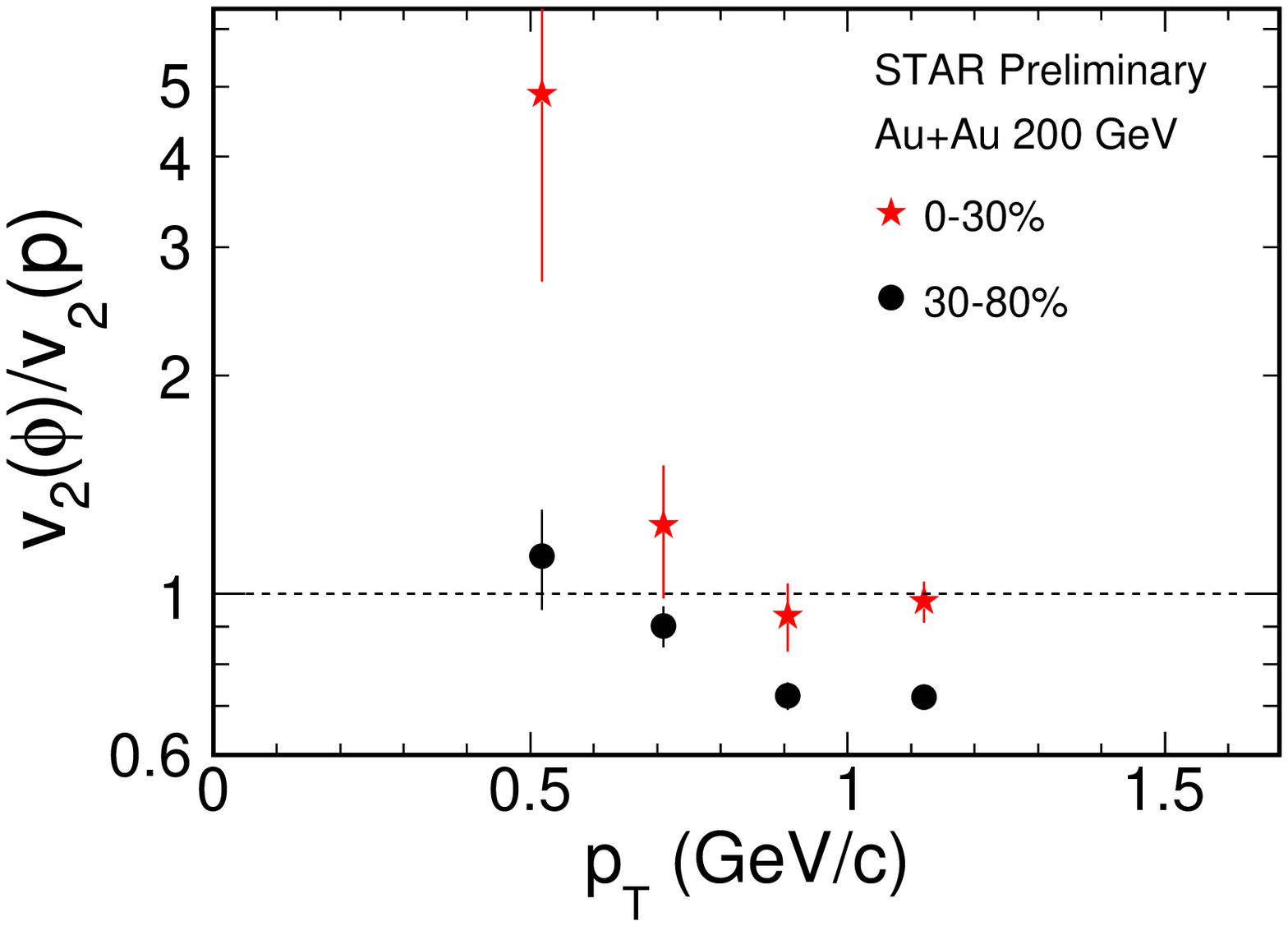}
\caption{Left: $R_{\rm{CP}}$ of $K^{*0}$ as a function of $p_T$ in Au+Au
  collisions at 200 GeV and 62.4 GeV compared with that of $K^0_S$ and
  $\Lambda$ at 200 GeV~\cite{Aggarwal:2010mt}. 
Right: $v_2(\phi)/v_2(p)$ ratio as a function of $p_T$ in Au+Au
collisions at two different centralities~\cite{fortheSTAR:2013cda}. 
} 
\label{rcpkstar_v2phi}       
\end{figure}
\begin{figure}
\centering
\sidecaption
\includegraphics[scale=0.3]{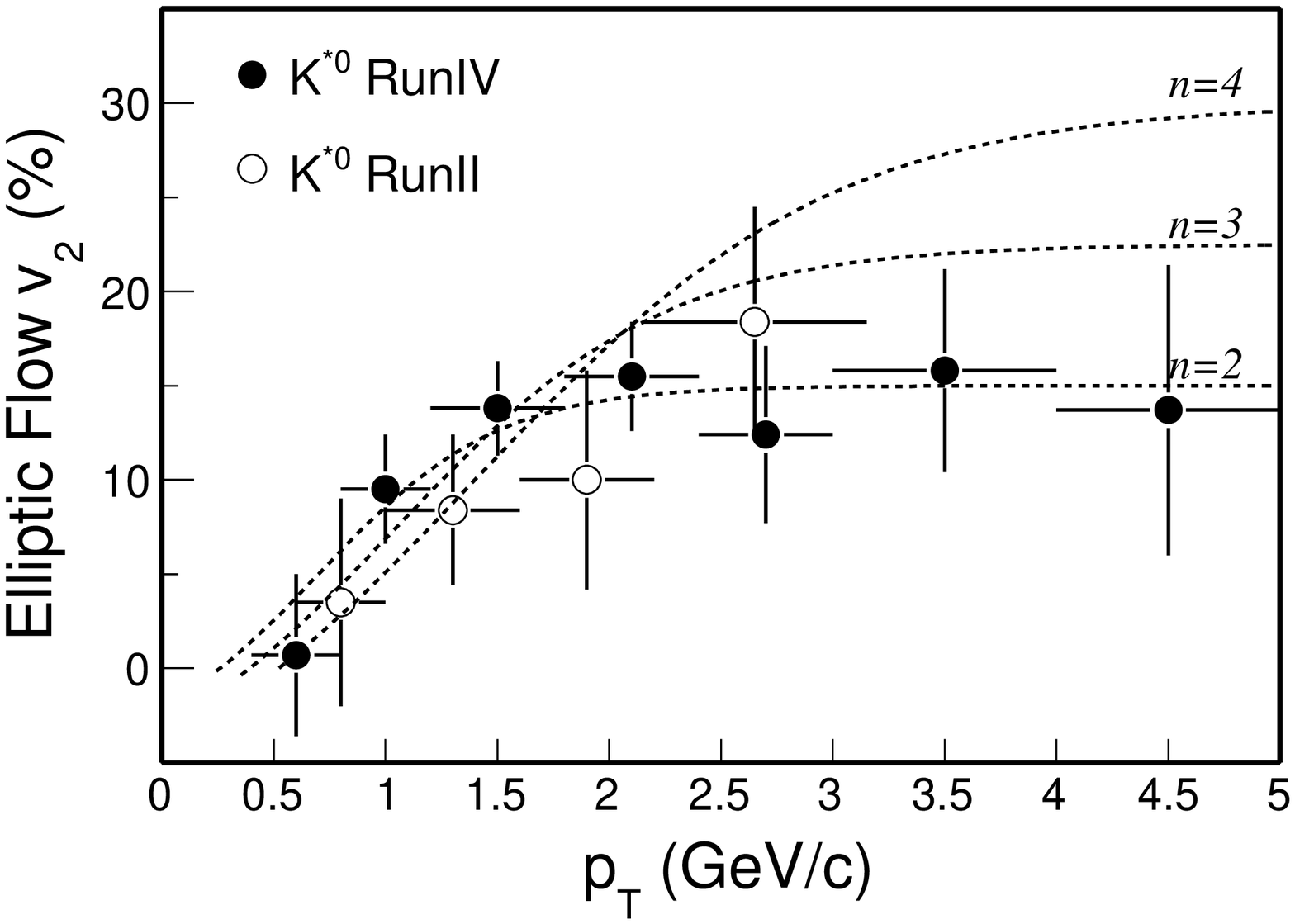}
 \includegraphics[scale=0.31]{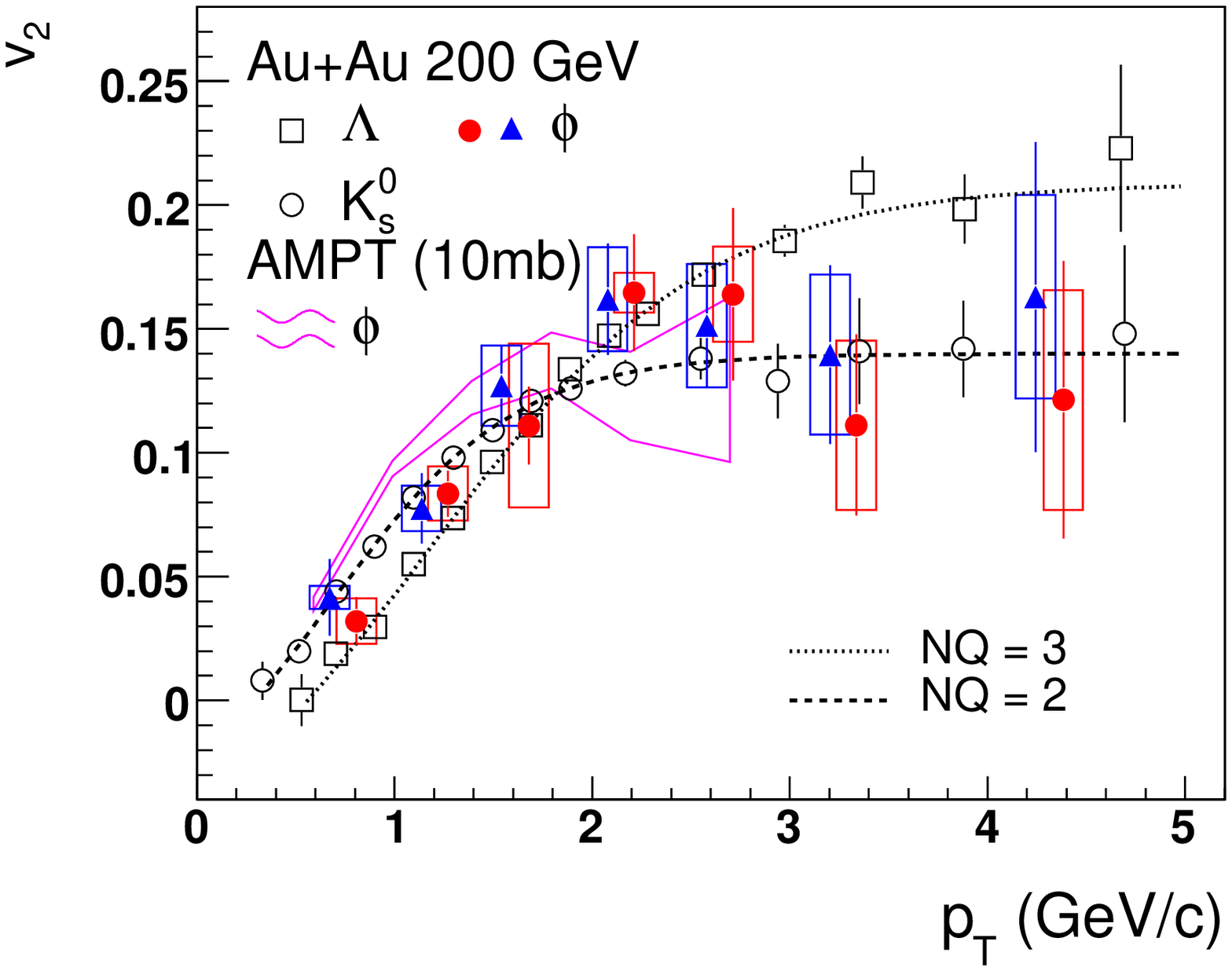}
\caption{Left: $K^{*0}$ $v_2$ as a function of $p_T$ in minimum bias Au+Au
  collisions at 200 GeV~\cite{Aggarwal:2010mt}. The dashed lines represent the $v_2$ of
  hadrons with different numbers of constituent quarks. Right: $p_T$
  dependence of $v_2$ of $\phi$, $\Lambda$, and $K^0_S$ in Au+Au
  collisions (0-80\%) at 200 GeV~\cite{Abelev:2007rw}. The magenta curved band represents
                  the $v_2$ of the $\phi$ meson from the AMPT model with a
                  string melting mechanism. The dash and dot
                  curves represent parametrizations for 
                  number of constituent quarks, NQ = 2 and NQ=3, respectively. }
\label{v2kstar_v2phi}       
\end{figure}
\begin{figure}
\centering
\sidecaption
\includegraphics[scale=0.28]{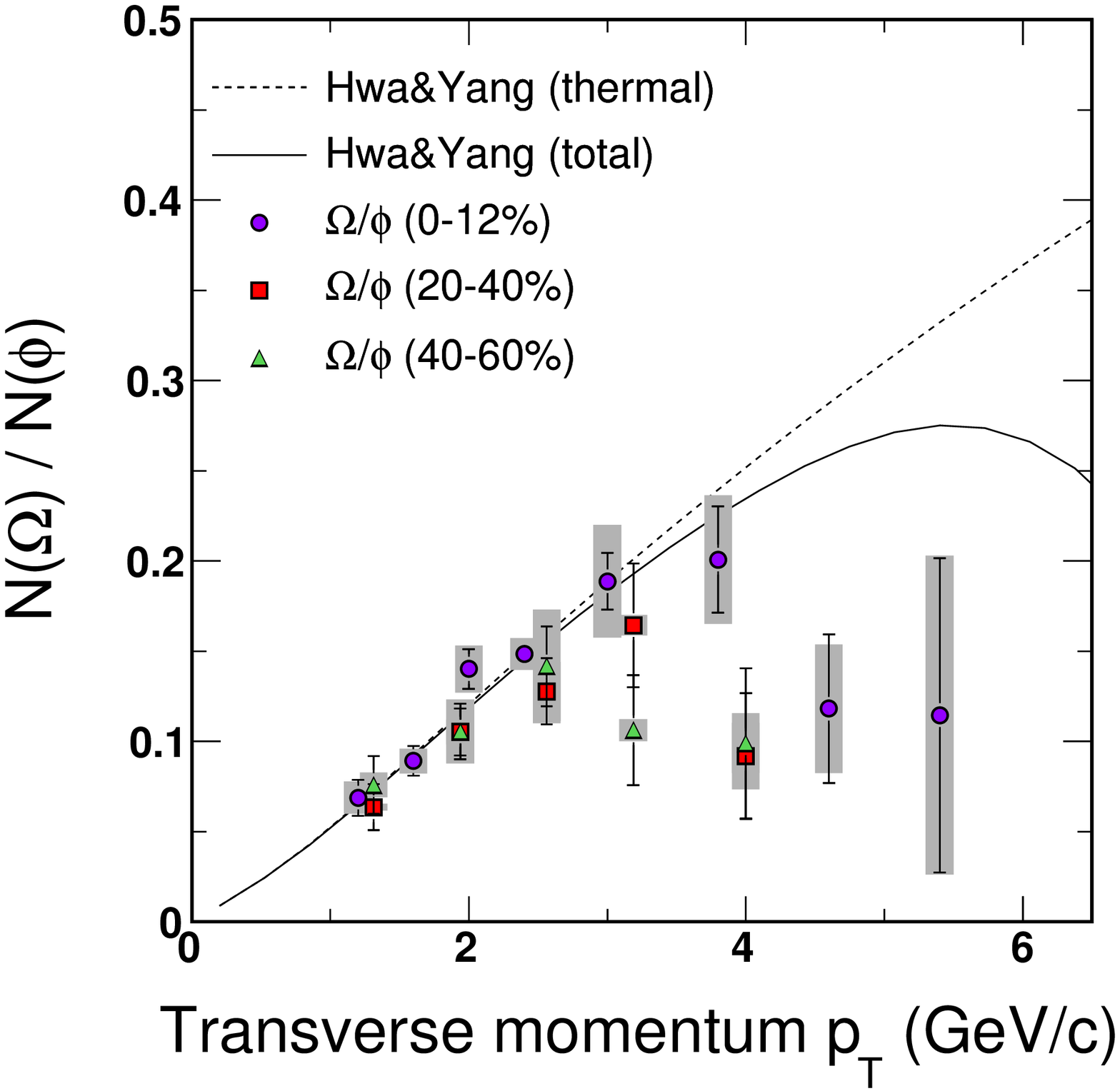}
\includegraphics[scale=0.28]{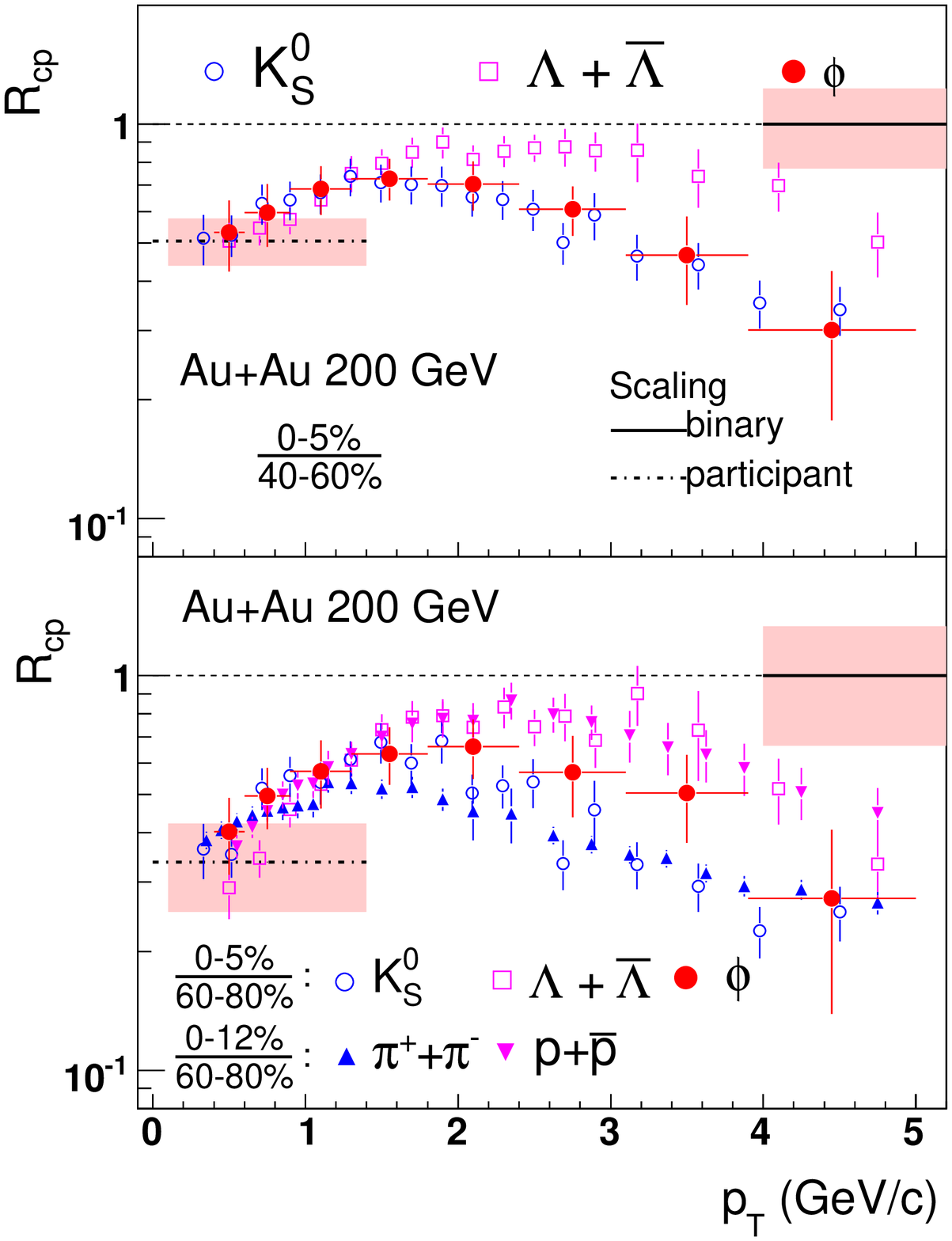}
\caption{Left: $\Omega/\phi$ ratio versus $p_T$ for three centralities
in $\sqrt{s_{NN}} =$ 200 GeV Au+Au collisions~\cite{Abelev:2007rw}.
The solid and dashed lines represent recombination model predictions
for central collisions for total and thermal contributions,
respectively. 
Right: $p_T$ dependence of the nuclear modification factor
$R_{\rm{CP}}$ $\phi$ meson compared with $K^0_S$ and $\Lambda$ in
Au+Au 200 GeV collisions~\cite{Abelev:2008aa}. 
The top and bottom
panels present $R_{\rm{CP}}$  from midperipheral and most-peripheral collisions,
respectively.
 }
\label{omegaphi_rcpphi}       
\end{figure}

\subsection{Rescattering effect}
Figure~\ref{ratio_kstar} (left plot) shows ratio of $K^{*0}$/$K^-$ as
a function of $dN_{\rm{ch}}/d\eta$, which reflects the
centrality~\cite{Abelev:2008yz}. The right plot shows the double ratio i.e. ratio of
$K^{*0}$/$K^-$ in heavy-ion collisions over that in $pp$ collisions~\cite{Aggarwal:2010mt}. One can see
that the ratio $K^{*0}$/$K^-$ decreases as a function of increasing
number of participating nucleons as well as decreases from $pp$,
$d$+Au, to central Au+Au collisions. This decrease in the
$K^{*0}$/$K^-$ ratio may be attributed to the rescattering of daughter
particles of $K^{*0}$.

Figure~\ref{ratio_phi} (left plot) shows ratio of $\phi$/$K^-$ as
a function of number of participating nucleons ~\cite{Abelev:2008aa}. The ratio remains flat
as a function of collision centrality, suggesting that there is a
negligible rescattering effect for $\phi$. The results are also
compared with UrQMD model which assumes kaon coalescence as the
dominant mechanism for $\phi$ production. As seen, the data rules out the kaon coalescence
as dominant mechanism for the $\phi$-meson production.
The right plot shows the double ratio i.e. ratio of
$\phi$/$K^{*0}$ in heavy-ion collisions over that in $pp$ collisions. 
It shows that the ratio increases with increasing
$N_{\rm{part}}$~\cite{Abelev:2008yz}. This increase might be either due to rescattering 
of daughter of $K^{*0}$ or strangeness enhancement for $\phi$.  Since,
we already see that $K^{*0}$ shows rescattering effect, this increase
is most likely due to the rescattering effect for $K^{*0}$.

Figure~\ref{rcpkstar_v2phi} (left plot) shows the nuclear
modification factor ($R_{\rm{CP}}$), defined as yields in central
collisions to that in peripheral collisions scaled by the number of
binary collisions~\cite{Aggarwal:2010mt,ref_rcp}. At low $p_T$ ($p_T < $1.8 GeV/c), we
observe that $R_{\rm{CP}}$ of $K^{*0}$ is less than that of $K^{0}_S$
(also a meson)
and $\Lambda$ (having almost similar mass). The
lower value of $R_{\rm{CP}}$ of $K^{*0}$ might be due to the
rescattering of daughter particles of $K^{*0}$ at low $p_T$ in the
medium.  The right plot shows the elliptic flow parameter $v_2$ of $\phi$
divided by $v_2$ of proton as a function of $p_T$ for high statistics Au+Au data at 200 GeV
for two different centralities~\cite{fortheSTAR:2013cda}. At low $p_T$, we observe that this
ratio is not unity. 
Since the $\phi$ mass is similar to the proton mass, we expect 
a similar $v_2$ for $\phi$ and proton
at low $p_T$ due to mass-ordering. 
However, data show that the mass ordering is broken at low $p_T$, which
might be due to rescattering of protons at low $p_T$ as suggested in Ref.~\cite{hydro_mod}.

\subsection{Quark coalescence and partonic effects}
Figure~\ref{v2kstar_v2phi} 
(left plot) shows the $v_2$ versus $p_T$ for $K^{*0}$ in Au+Au
collisions at 200 GeV~\cite{Aggarwal:2010mt}. The various
curves show 
the number of constituent quarks ($n=$2
for mesons and 3 for baryons). We observe 
that the $K^{*0}$ $v_2$ follows the $n=$2 parametrization suggesting quark coalescence
for their production~\cite{quark_coal}. 
The right plot shows the $v_2$ versus $p_T$ for the $\phi$ meson compared with $K^0_S$ and $\Lambda$
along with number-of-constituent-quark
parameterizations~\cite{Abelev:2007rw}. We see that the 
$\phi$ meson follows the $K^0_S$ behavior and the $n=$2 curve. 
Since $\phi$ meson is not formed via kaon
coalescence and undergoes less hadronic interaction (as discussed
before), the observed $v_2$ of $\phi$ is due to the partonic phase. 
These results also suggest that heavier quarks flow as strongly
as lighter quarks. 

Figure~\ref{omegaphi_rcpphi} (left plot) shows the $\Omega$/$\phi$
ratio as a function of transverse momentum $p_T$ for three different
centralities in Au+Au collisions at 200 GeV~\cite{Abelev:2007rw}.  Data are compared with 
the model calculations which assume that $\Omega$ and $\phi$ are 
produced from thermal $s$ quarks coalescence
in the medium. The results show that the coalescence model reproduces 
the data at low $p_T$. The right plot shows the nuclear modification factor $R_{\rm{CP}}$ of $\phi$
for 0--5\%/40-60\% (top panel) and 0--5\%/60-80\% (bottom panel),
compared with $R_{\rm{CP}}$ of different
particles~\cite{Abelev:2008aa}. A suppression of
$R_{\rm{CP}}$ at high $p_T$ has been suggested to be the signature of
dense medium or quark gluon plasma formation in heavy-ion
collisions. Together with Fig.~\ref{rcpkstar_v2phi} (left plot), above
results suggest that both $K^{*0}$ and $\phi$ $R_{\rm{CP}}$ show
suppression at high $p_T$, suggesting dense medium formation at the top
RHIC energy. 

\begin{figure}
\centering
\sidecaption
\includegraphics[scale=0.45]{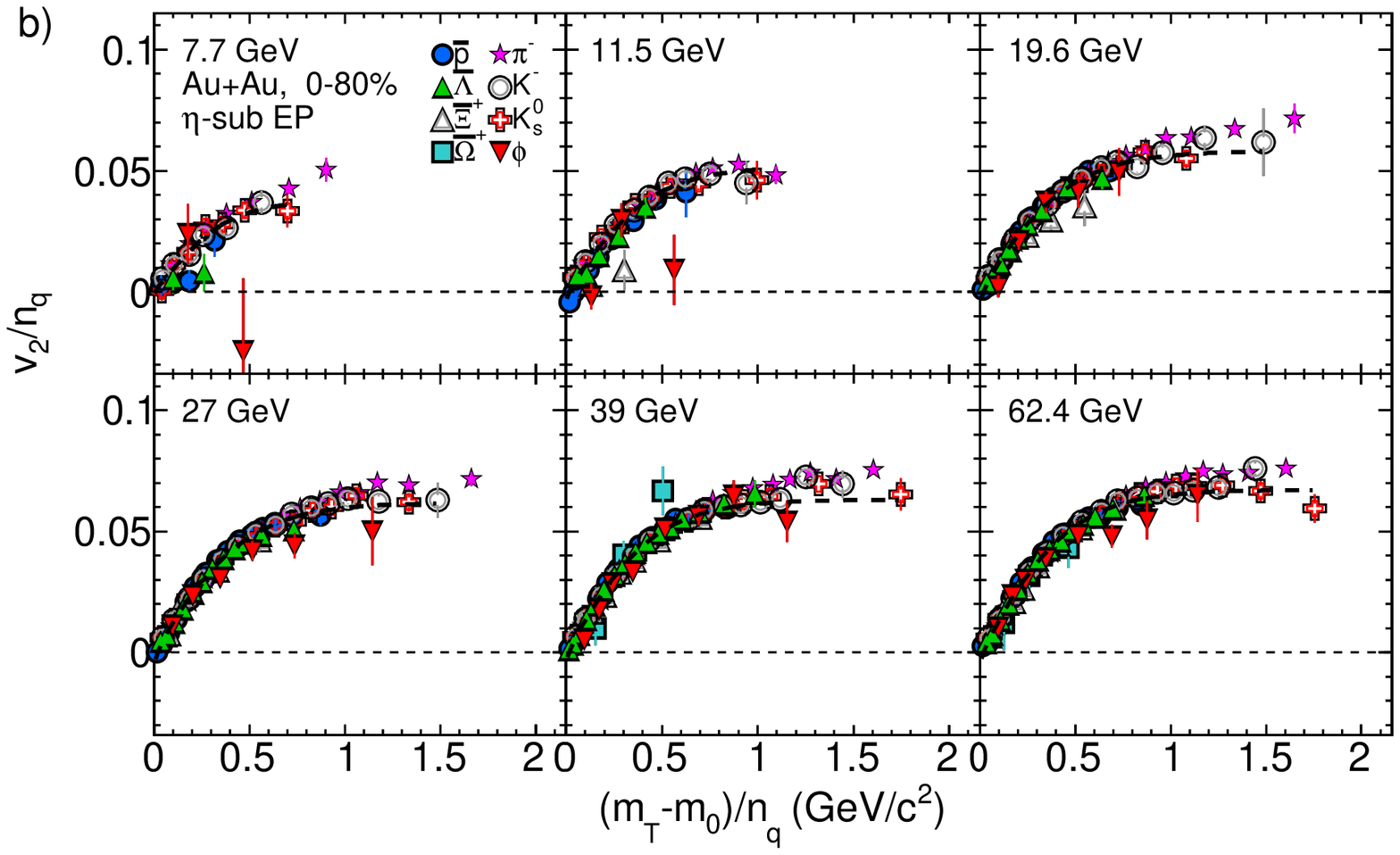}
\includegraphics[scale=0.26]{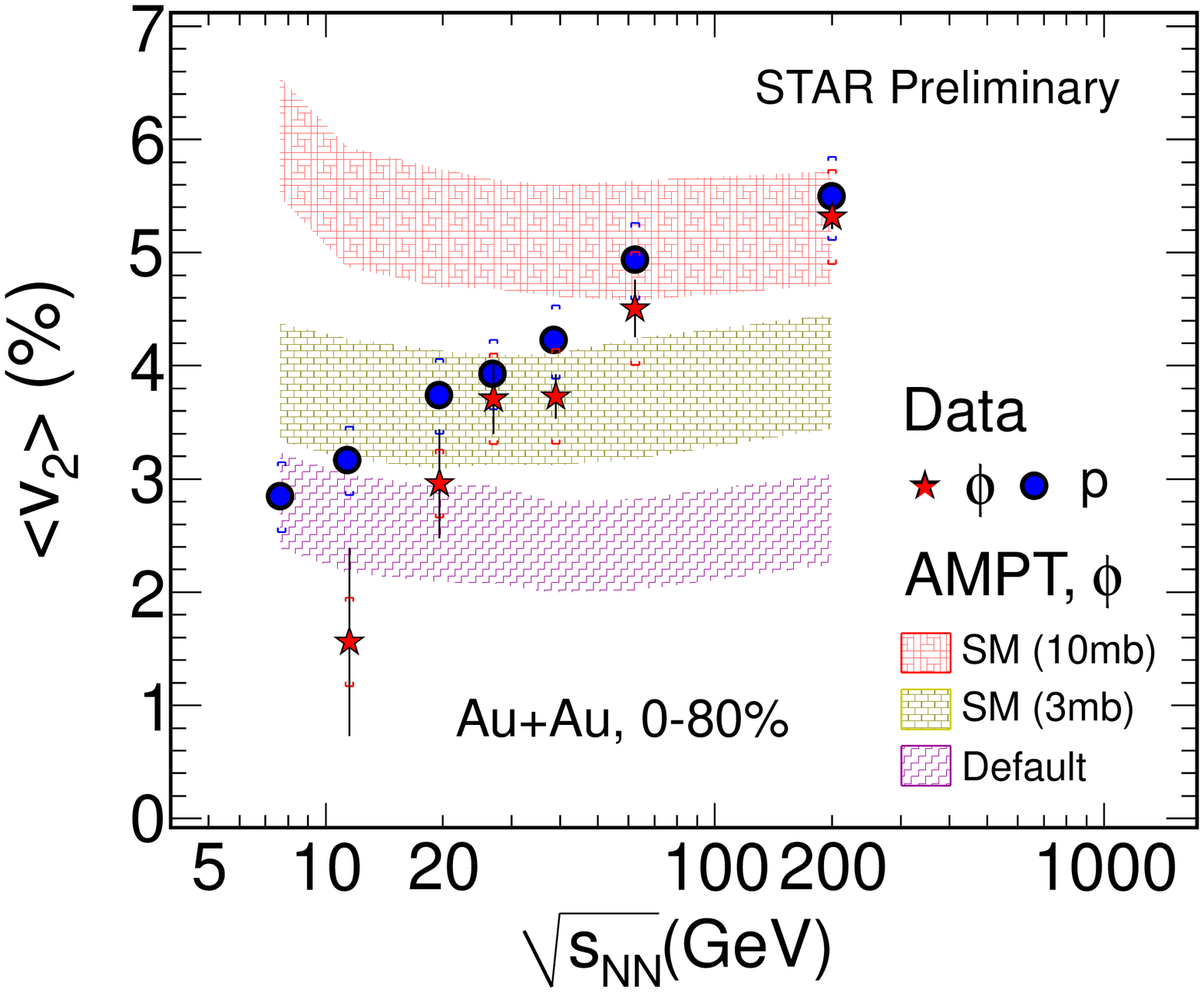}
\caption{Left: $v_2/n_q$ as a function of $(m_T-m_0)/n_q$
 for
  different
particles in Au+Au collisions at $\sqrt{s_{NN}} =$ 7.7, 11.5, 19.6,
27, 39 and 62.4 GeV~\cite{v2_prl_bes}.
Right: The $p_T$ integrated $\phi$ meson and proton $v_2$ for Au+Au
minimum bias  (0–-80\%) collisions at mid-rapidity $|y|<$ 1.0 at RHIC
as a function of  $\sqrt{s_{NN}}$~\cite{ref_wp}. 
The $\phi$ meson $v_2$ values are compared with corresponding AMPT model calculations at various beam energies.
}
\label{v2bes}      
\end{figure}
\begin{figure}
\centering
\sidecaption
 \includegraphics[scale=0.32]{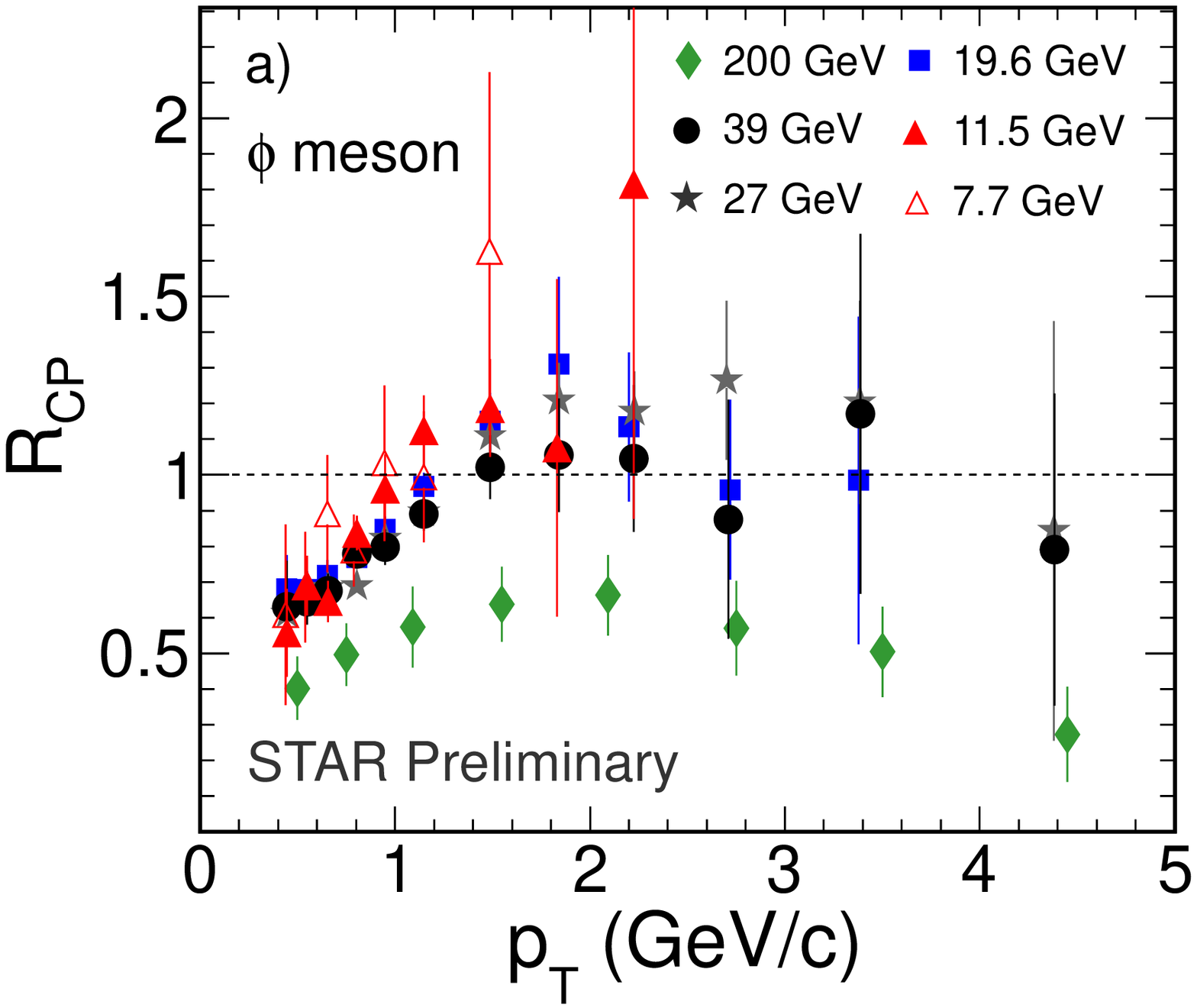}
\includegraphics[scale=0.28]{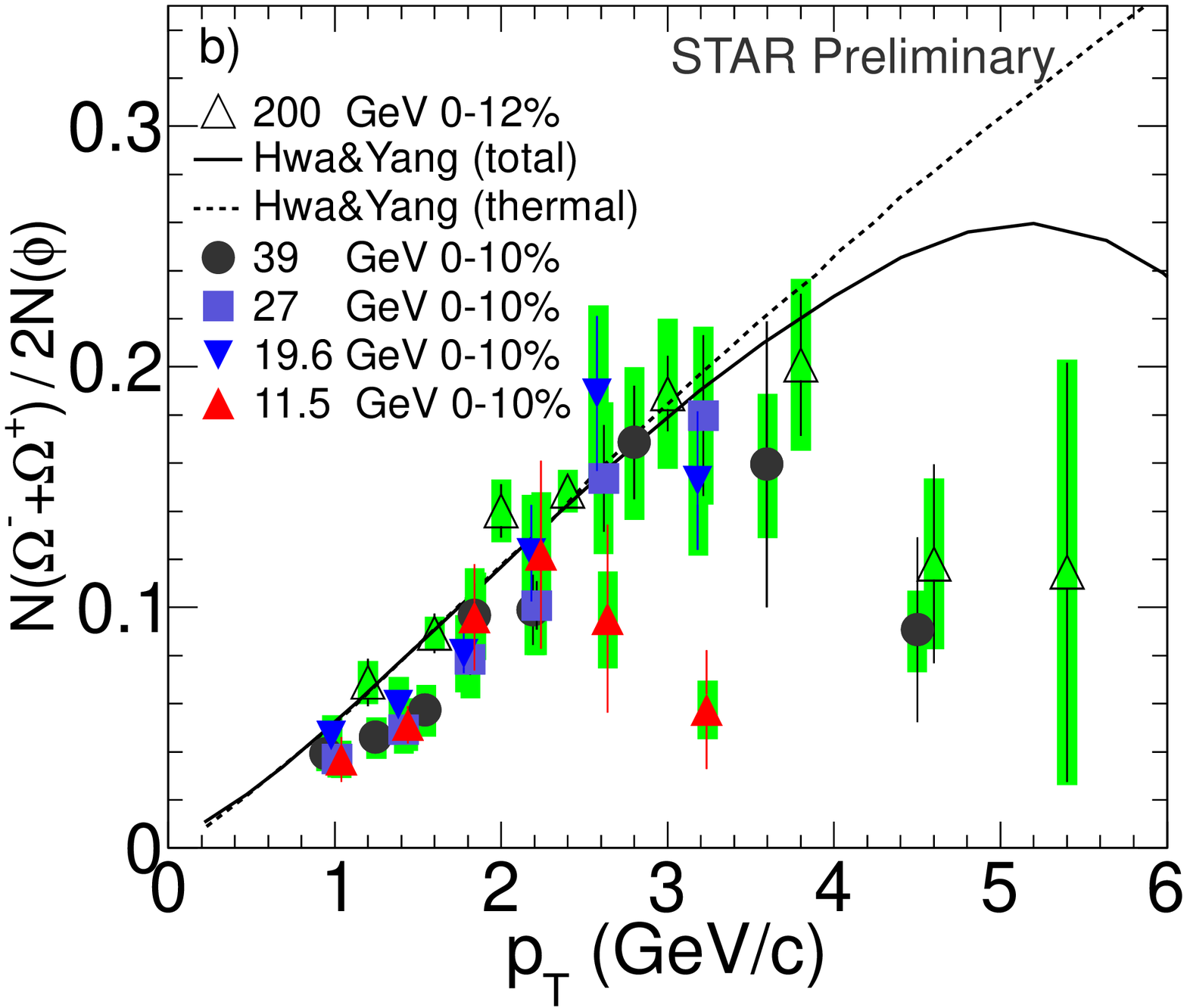}
\caption{Left: $R_{\rm{CP}}$ of
$\phi$ as a function of $p_T$ in Au+Au collisions at various beam energies~\cite{fortheSTAR:2013cda}.
Right: Ratio N($\Omega^- + \Omega^+ )/(2N \phi$) as a function of $p_T$ in central Au+Au collisions at
$\sqrt{s_{NN}}=$11.5–200 GeV~\cite{ref_wp,mpla_lok,fortheSTAR:2013gwa}. The curves represent model calculations
by Hwa and Yang for $\sqrt{s_{NN}} =$ 200 GeV.}
\label{omegaphi_rcp_bes}       
\end{figure}
\subsection{Energy dependence of partonic interactions}
In the previous subsection, we have established the partonic nature of the
system formed at the top RHIC energy. It is interesting to see what happens to this partonic nature
when the collision energy is decreased. RHIC Beam Energy Scan (BES)
allows to check this energy dependence. 

Figure~\ref{v2bes} (left plot) shows the 
$v_2$ scaled by the number of constituent quarks (ncq)
plotted versus $m_T$-$m$ divided by number of constituent 
quarks. Results are shown for 7.7, 11.5, 19.6, 27, 39, and 62.4 GeV, and for various
particles that include mesons and baryons~\cite{v2_prl_bes}. We observe that all particles
follow ncq scaling down to 19.6 GeV. However, at 11.5 GeV and below 
$\phi$ mesons deviate from this scaling. Since $\phi$ meson has a
small  hadronic interaction cross-section~\cite{Adams:2005dq}, this small $\phi$ $v_2$ may suggest
less partonic contributions at lower energies. However, as can been
seen, higher statistics are needed at lower energies to make definite conclusions.
The right plot shows the the $v_2$ of $\phi$ mesons compared to corresponding AMPT model
calculations~\cite{ref_wp}. The $\langle v_2 \rangle$
values from the model is constant for all the energies at a given parton-parton interaction cross-section. 
This is expected because it is the interactions between minijet partons in the AMPT models that generate $v_2$. 
The $v_2$ of $\phi$ mesons for $\sqrt{s_{NN}} >$ 19.6 GeV can be explained by the AMPT
model with string melting enabled (AMPT-SM). The AMPT-SM 
model with 10 mb parton-parton cross-section fits the data at
$\sqrt{s_{NN}} =$ 62.4 and 200 GeV, whereas a reduced  value of
parton-parton cross-section of 3 mb is needed to describe  
the data at  $\sqrt{s_{NN}} =$ 27 and 39 GeV. On the other hand, the data at
$\sqrt{s_{NN}} =$ 11.5 GeV are explained  within the default version
of the  AMPT model without the partonic interactions. These model
results along with data indicate that for $\sqrt{s_{NN}} <$ 11.5 GeV, the hadronic
interaction plays a dominant role, whereas above 19.6 GeV contribution
from partonic interactions increases.

Figure~\ref{omegaphi_rcp_bes} (left plot) Nuclear modification factor $R_{\rm{CP}}$
(0--10\%/40--60\%) of $\phi$ meson at
different BES energies along with 200 GeV ((0--5\%/40--60\%)~\cite{fortheSTAR:2013cda}. We observe that $R_{\rm{CP}} \ge $ 1 for beam
energies  $\sqrt{s_{NN}} \le$ 19.6 GeV. 
The right plot shows the $\Omega$/$\phi$
ratio versus $p_T$ for different energies from $\sqrt{s_{NN}} =$ 11.5, 19.6 GeV, up to 
200 GeV~\cite{ref_wp,mpla_lok,fortheSTAR:2013gwa}. We observe that 19.6 , 27 and 39 GeV follow the same behavior
as  200 GeV, however, the ratio at 11.5 GeV show different trend
i.e. the ratio turns down at lower $p_T$ when compared to higher energies. This may suggest different particle 
production phenomenon at 11.5 GeV compared to higher energies.

\section{Summary}
In summary, $K^{*0}$(892) and $\phi$(1020) resonance production at
RHIC is discussed. 
The $K^{*0}$/$K^-$ ratio decreases as a function of
$N_{\rm{part}}$, $R_{\rm{CP}}$ of $K^{*0}$ is less than that of $K^{0}_S$
and $\Lambda$ at low
$p_T$, and the $\phi/K$ ratio remains constant as a function of
centrality. These results suggest rescattering effect for $K^{*0}$
and may be negligible rescattering effect for $\phi$ mesons. 
The number-of-constituent-quark scaling is observed for both $K^{*0}$ and
$\phi$, possibly indicating partonic nature of system formed at the top RHIC
energy. Similarly, the quark coalescence model explains the $\Omega$/$\phi$ ratio
at low $p_T$.
The energy dependence of various observables suggest that the system formed at
lower energies may be hadron dominant. As an example, $\phi$ meson
$v_2$ does not follow ncq-scaling for $\sqrt{s_{NN}} \le$ 11.5
GeV. $v_2$ of $\phi$ mesons compared with model
results indicate that for $\sqrt{s_{NN}} <$ 11.5 GeV the hadronic
interaction plays a dominant role. Energy dependence of
$\Omega$/$\phi$ ratio versus $p_T$ suggests
a change of particle production at $\sqrt{s_{NN}} =$ 11.5 GeV and the
$\phi$ meson $R_{\rm{CP}} \ge $ 1 for beam energies  $\sqrt{s_{NN}} \le$ 19.6 GeV. 

%
%
%

\end{document}